# Structural response of α-quartz under plate-impact shock compression

Sally June Tracy[1,2]*, Stefan J. Turneaure[3], and Thomas S. Duffy[1]

[1] Department of Geosciences, Princeton University, Princeton, New Jersey 08544, USA.
[2] Geophysical Laboratory, Carnegie Institution for Science, Washington, DC 20015, USA.
[3] Institute for Shock Physics, Washington State University, Pullman, Washington 99164, USA.

*To whom correspondence should be addressed; E-mail: sjtracy@carnegiescience.edu

**Abstract**

Due to its far-reaching applications in geophysics and materials science, quartz has been one of the most extensively examined materials under dynamic compression. Despite 50 years of active research, questions remain concerning the structure and transformation of $SiO_2$ under shock compression. Continuum gas-gun studies have established that under shock loading quartz transforms through an assumed mixed-phase region to a dense high-pressure phase. While it has been often assumed that this high-pressure phase corresponds to the stishovite structure observed in static experiments, there has been no atomic-level structure data confirming this. In this study, we use gas-gun shock compression coupled with *in-situ* synchrotron X-ray diffraction to interrogate the crystal structure in shock-compressed α-quartz up to 65 GPa. Our results reveal that α-quartz undergoes a phase transformation to a disordered metastable phase as opposed to crystalline stishovite or an amorphous phase, challenging long-standing assumptions about the dynamic response of this fundamental material.

**Introduction**

Laboratory shock wave experiments have long played an important role in characterizing properties of geophysical materials at the high pressure and temperature conditions of the deep Earth (*1*). Shock-compression experiments yield pressure-temperature states comparable to planetary adiabats, thus requiring minimal extrapolation for geophysical application. Moreover, shock loading presents a unique capability to study impact phenomena in real time, providing insight into natural impact processes relevant to planetary formation and evolution. Quartz ($SiO_2$) is one of the most abundant minerals of Earth's crust and is widely distributed in different rock types. As a result, characterizing the dynamic response of $SiO_2$ is important for interpretation of shock metamorphism in samples from terrestrial impact sites (*2, 3*). Furthermore, quartz serves as an archetype for the silicate minerals of the crust and mantle. As such, characterizing high-pressure, high-temperature behavior of quartz is important for understanding potential silica-rich regions of the deep Earth.

The high-pressure behavior of quartz has been the subject of extensive experimental studies. At ambient conditions, the stable polymorph of $SiO_2$ is the trigonal α-quartz structure ($P3_221$), composed of a corner-linked framework of $SiO_4$ tetrahedra. Based on the equilibrium phase diagram, under pressure $SiO_2$ adopts a series of crystalline phases from quartz to coesite ($C2/c$) to stishovite (rutile structure, $P4_2/mnm$) to the $CaCl_2$-type phase ($Pnnm$) (*4*). In addition, a number of metastable forms of $SiO_2$ have been observed experimentally or predicted theoretically (*5–10*). Due to the high kinetic barriers associated with these transitions, when compressed in a diamond anvil cell (DAC) at 300 K, quartz persists to above 20 GPa where it transforms to a disordered, dense metastable phase (*7, 11*).

The behavior of quartz under shock compression differs markedly from its static response. The Hugoniot (Fig. 1) is characterized by a highly compressible region often called the "mixed-phase region" that initiates at ~15 GPa and reaches completion at ~40 GPa, at which point the material compressibility decreases significantly. Based on impedance matching Hugoniot data and thermodynamic considerations, it is often assumed that the high-pressure phase on the quartz Hugoniot corresponds to crystalline stishovite (*12–14*). However, there is no direct evidence demonstrating this and the structure of the high-pressure phase remains a subject of continued debate.

The sequence of phase transitions occurring in dynamically compressed quartz and the role of kinetics remain unknown. For $\alpha$-quartz shocked between 30 and 65 GPa, calculated Hugoniot temperatures range from ~1500-4000 K (*15*). Recovery experiments and samples collected at natural impact sites are found to consist primarily of amorphous material with trace quantities of stishovite (*16*). Due to the high-temperature release path, it is an open question whether amorphization occurs during compression or release. On more general grounds, it has been suggested that reconstructive transformations involving tetrahedral to octahedral coordination changes are kinetically limited on shockwave timescales and hence there is insufficient time for formation of stishovite in a laboratory shockwave experiment (*17*). Accordingly, it has been contended that the high-pressure phase of quartz instead corresponds to a metastable intermediate or a dense amorphous structure (*18–20*). Using traditional continuum diagnostics, the crystal structure of high-pressure phases formed under dynamic compression cannot be determined experimentally. As a result, key questions concerning the nature of the mixed-phase region and the structure of the high-pressure phase(s) on the Hugoniot remain unresolved.

Recently, *in situ* X-ray diffraction measurements under gas-gun loading have shown that fused silica remains amorphous under shock compression until 35 GPa and for higher stresses it transforms to polycrystalline stishovite (*21*). Stishovite formation has also been observed in laser-shock experiments in fused silica (*22*) and in recovery experiments on porous sandstone (*23*). Here we use time-resolved X-ray diffraction measurements coupled with gun-based dynamic compression to probe the crystal structure of $\alpha$-quartz under shock compression. Plate impact loading provides a uniform, well-defined state of uniaxial strain within the shocked quartz. The X-ray diffraction data provide a complete picture of the material response by revealing new details of atomic level structure for stresses from 31-65 GPa.

**Results**

Figure 2 shows representative two-dimensional (2D) X-ray diffraction (XRD) images from Z-cut quartz shots at 31 and ~63 GPa. For the lower stress experiment (Fig. 2a), peaks corresponding to compressed $\alpha$-quartz can be identified. In the 2D image, the compressed peaks appear as localized bright spots much broader than the ambient single crystal Laue spots, indicating that the shocked Z-cut quartz developed a mosaic spread of several degrees (Fig. 2a). Using the intense (101) peak, the density of the compressed quartz can be estimated to be ~3.65 g/cm$^3$. This density is consistent with an extrapolation of the 300-K equation of state of $\alpha$-quartz from static compression experiments (*24*), where the modest offset is consistent with the higher temperature along the Hugoniot (Fig. 1). In addition to the α-quartz peaks, there are a number of other peaks that cannot be assigned to the α-quartz structure indicating that the material has partially undergone a pressure-induced phase transformation. The combination of low- and high-pressure phase material indicates that the Hugoniot state is a mixed-phase region near 30 GPa. The scattering from the transformed material has more extended arcs, albeit with pronounced azimuthal intensity variations. This finding indicates that preferred orientation (or texture) persists above the phase transformation.

At ~63 GPa (Fig. 2b), the α-quartz peaks have disappeared, indicating complete transformation to a high-pressure phase. For the higher angle peaks, the width of the profiles indicates modest broadening on top of the expected instrumental contribution. These results demonstrate that quartz transforms to a high-pressure phase with crystalline order and clearly shows that the high-pressure phase is not amorphous as had been proposed in the literature (*18*, *19*, *25*).

Figure 3 shows a series of azimuthally integrated diffraction patterns for Z-cut quartz obtained for stress states ranging from 31 to 65 GPa. Other than the XRD profile collected at 65 GPa, all measured XRD profiles shown in Fig. 3 are from frames captured prior to the shock wave having traversed the entire thickness of the Z-cut quartz sample and correspond to Hugoniot states. The pattern collected at 65 GPa corresponds to a double-shock state produced by re-shock from 56 to 65 GPa at the Lithium Fluoride (LiF) window. Shocked Z-cut quartz diffraction data collected for Hugoniot stresses at 31-35 GPa exhibit an intense peak at ~10° two-theta, consistent with the compressed α-quartz (101) reflection. Additionally, new diffraction peaks are observed at 16° and 21°. For shock stresses above 35 GPa, the α-quartz (101) peak can no longer be detected and the diffraction pattern contains a broad feature spanning 9-12° two-theta with distinctly sharper peaks near 16° and 21°. These two new relatively sharp peaks are from discrete regions around the diffraction rings (see Fig. 2) indicating significant texture both in the mixed-phase region and in the high-pressure region.

Figure 4 shows a comparison of diffraction data from shocked X-cut, Y-cut, and Z-cut quartz collected at 56-65 GPa. Similar to the Z-cut quartz results, the diffraction patterns for X-cut and Y-cut quartz also indicate significant texture of the high-pressure phase as evidenced by the observed large intensity variations around the diffraction rings While the azimuthal intensity variations of the diffraction patterns are different for different orientations, the integrated patterns indicate that all orientations transform to the same phase under shock compression.

Figure 5 compares the diffraction data for the shocked polycrystalline novaculite (natural microcrystalline quartz) with the diffraction patterns collected for Z-cut quartz and fused silica shocked to similar peak stresses (63-66 GPa). In previous experiments at the DCS, we used *in situ* XRD coupled with gun-based shock compression to examine the structure of shocked fused silica (*21*). That study demonstrated that silica glass adopts a dense amorphous structure for shock pressures up to 35 GPa, above which it transforms to nanocrystalline stishovite. A comparison of the stishovite diffraction pattern obtained from fused silica starting material to the diffraction patterns for α-quartz starting material show that while the integrated diffraction patterns for single-crystal and polycrystalline quartz have similarities they are distinct from the stishovite pattern observed for shocked fused silica.

The three shots in Fig. 5 all correspond to double shock measurements, where XRD data were collected after the shock waves propagated through the SiO$_2$ sample reflecting from the LiF window as a reshock resulting in a uniform final stress state. Diffraction data for polycrystalline novaculite samples shocked to lower peak stresses show similar features (see Supplemental Material, Fig. S7). In the polycrystalline data, the strong peak at 9.5 degrees corresponds to a residual ghost peak from the ambient strong (101) α-quartz reflection. This peak is the result of incomplete decay of the phosphor scintillator detector between successive X-ray frames and does not arise from scattering from the compressed sample. The signal intensity is notably lower for the polycrystalline data compared to the single-crystal data, yielding lower signal-to-noise. While the integrated diffraction pattern is similar to single-crystal data, the diffraction rings are smooth and continuous without the azimuthal intensity variations from preferred orientation observed in the single-crystal data.

**Discussion**

A comparison of the diffraction patterns for the high-pressure phases of silica glass and α-quartz (Fig. 5) at similar shock pressures highlights the clear distinction in the structure of the high-pressure phase depending on starting material. Here, we reveal two key distinctions in the structure of α-quartz under shock loading compared with fused silica: (1) As opposed to the densification observed below 35 GPa for silica glass, the region on the α-quartz Hugoniot between ~30-40 GPa represents a crystalline mixed-phase region, that is, a coexistence of compressed α-quartz and transformed high-pressure material; (2) while α-quartz undergoes a phase transformation to a crystalline phase at a similar shock-pressure as silica glass, the structure of the high-pressure phase bears similarity to the stishovite structure but is distinctly different. In particular, the diffraction pattern from α-quartz does not exhibit an intense low-angle peak corresponding to the stishovite (110) reflection.

The absence of the strong stishovite (110) peak is consistent across multiple orientations of single crystals (Fig. 4) as well as polycrystalline starting material (Fig. 5) and therefore cannot be attributed to residual texture due to an orientation relationship between the starting material and the high-pressure phase. While we do not observe a sharp low-angle peak in the XRD patterns, there is a broad, weaker feature peaked at Q~2.3 Å$^{-1}$ [Q=4πsin(θ)/λ] that persists across the entire measured stress range (Fig. 3). One possible explanation is that this feature corresponds to partial amorphization of the sample (*16–20, 26, 27*). However, this feature exhibits some azimuthal intensity variation on the 2D diffraction image, inconsistent with scattering from an amorphous material. Furthermore, this low-angle feature occurs at lower Q than the first strong diffraction peak (FSDP) of SiO$_2$ glass observed in static compression experiments in this pressure range (see Fig. S7) (*28–31*). This differs from shocked silica glass (*21*), where the FSDP overlaps static data. While the higher temperature of the Hugoniot states needs to be accounted for, recent results show that elevated temperatures can enable additional compression mechanisms, allowing the glass to achieve a denser state than it can at lower temperatures (*32, 33*). Additionally, the broad character and low intensity relative to other peaks remain fairly constant with increasing pressure, despite the large change in expected temperature over this range which should promote crystallization.

The textured Debye-Scherrer rings of the high-pressure phase (Fig. 2) are in marked contrast with the results of recent molecular dynamics (MD) simulations for both α-quartz and silica glass (*34*). In these simulations, uniaxial loading induces picosecond amorphization of the α-quartz crystal structure, followed by nanosecond stishovite crystallization via a homogeneous nucleation and growth. After amorphization, the simulations for α-quartz proceed via a transformation mechanism that mirrors that of silica glass under similar loading conditions. Unlike silica glass, where we observed smooth powder-like rings (*21*), the residual texture in our data is hard to reconcile with a nucleation and growth model. It is unlikely that within the timescale of our measurements (several hundred nanoseconds) the requisite grain growth could occur to yield this degree of texture. Instead, the orientation-dependent texture suggests a topotactic relationship between the α-quartz starting material and the high-pressure phase. Accordingly, it is likely the transformation involves a displacive or shear mediated mechanism preserving some degree of atomistic neighbor memories.

Figure 1 shows the Hugoniot for α-quartz in pressure-volume space including selected previous continuum data (*35–38*). Also shown are 300-K static compression data for stishovite (*39*) along with the densities determined from fits to our XRD data in the high-pressure region. Details of the fits including the accounting for the spectral shape of the pink X-ray beam are described in detail in the supplemental material. X-ray densities derived from fits assuming the stishovite structure are denser than both the continuum Hugoniot as well as the 300-K isothermal data for stishovite. The high temperatures generated during shock loading require the Hugoniot necessarily be offset to lower densities relative to the 300-K isotherm. As such, the assignment of the high-pressure phase under shock loading to the stishovite structure can be ruled out.

Numerous metastable high-pressure polymorphs of $SiO_2$ have been reported both from theoretical calculations and static compression experiments (*5–10, 40–42*). These structures can generally be described as various silicon cation fillings of the octahedral voids within an approximately hexagonal-close packed (hcp) oxygen lattice (*6*). A wide range of energetically competitive structures can be generated by modifying the silicon filling pattern and the degree of distortion of the oxygen sublattice. Within this context, stishovite can be described as a distorted hcp array of oxygen anions where one-half of the available octahedral interstices are filled by silicon ions forming linear chains of edge-sharing $SiO_6$ octahedra. When transposed into the rutile unit cell, this silicon ordering represents a filling of every other void along the tetragonal <110> direction, yielding an intense (110) peak in the stishovite diffraction pattern. As a result, any disorder in the silicon sublattice causes destructive interference along this direction, effectively reducing the intensity of the (110) peak.

The defective niccolite structure (*d*-NiAs or $Fe_2N$-type) can be considered the most general structure within this family of $SiO_2$-oxygen close-packed phases. In this structure, the oxygen anions are arranged in an ideal hcp lattice and silicon cations are distributed randomly across the octahedral voids with an occupancy factor of ½. This structure has been observed as a metastable phase in heated diamond anvil experiments for pressures between 30-60 GPa and temperatures between 900-1200 K (*8, 41, 43*). While the overwhelming majority of the many gas-gun shock recovery experiments on $SiO_2$ report finding quartz or amorphous material, there are two reports in which a very small amount of *d*-NiAs phase was identified together with quartz and glass (*40, 44*). Figure 3 includes simulated diffraction patterns for both the *d*-NiAs and stishovite structures.

While the patterns are generally similar, the stishovite pattern contains additional peaks that arise from coherent scattering from ordered planes of silicon within the structure. Notably, the $d$-NiAs pattern has no low-angle peak due to the lack of any long-range silicon order. Furthermore, for structures of equivalent density, the stishovite peaks are offset to slightly lower angles. This shift arises from the lower packing efficiency of the distorted oxygen framework as opposed to the ideal hcp lattice of the $d$-NiAs structure. In addition to stishovite densities, Fig. 1 also includes densities derived from fits to the $d$-NiAs structure. In comparison to the unphysically high densities determined from the stishovite fits, Le Bail profile refinements assuming the $d$-NiAs structure yield densities that are consistent with the continuum Hugoniot data. Fits to diffraction data for X- and Y-cut quartz starting material as well as novaculite yield an overall similar result (Supplemental Material, Fig. S6). The densities in the mixed-phase region could not be accurately determined due to overlapping diffraction peaks for multiple phases.

While fits to the $d$-NiAs structure agree with the previous continuum pressure-volume Hugoniot results (35–38), this structure cannot account for the diffuse low-angle peak observed consistently across the high-pressure region (Fig. 3). It is possible to explain the XRD patterns in terms of $d$-NiAs phase coexisting with some amount of dense amorphous material. In this case, the diffuse peak could be attributed to retention of some amount of an amorphous metastable intermediate, consistent with molecular dynamics simulations (34). Although, the azimuthal intensity variation (Fig. 2b) as well as the discrepancy in the position of this feature compared to the FSDP from $SiO_2$ glass under static compression complicates this interpretation.

A second explanation for the observed diffraction patterns in the high-pressure region is a structure composed of a well-defined hcp oxygen framework filled with silicon cations lacking well-defined long-range order. A comparison of diffraction patterns for the various metastable high-pressure polymorphs of $SiO_2$ reveals an overall similarity in the high scattering angle peaks yet distinctions in the low-angle peak position(s) (Supplemental Material, Fig. S8). These differences can be attributed primarily to various silicon ordering schemes within an hcp-like oxygen lattice. The diversity of metastable $SiO_2$ phases indicates a complex energy landscape with numerous competing structures of similar energy.

There is a large body of work exploring both the structural interrelationships and transformation mechanisms between low-pressure $SiO_2$ tetrahedral-coordination structures and various high-pressure six-coordinated phases, both stable and metastable (6, 45–47). Within this context, each structure is defined by an ordering of the silicon cations within the tetrahedral and octahedral voids in a quasi-close-packed oxygen framework. The pathways between structures can be broken down into sets of transformations involving some combination of displacive and ordering mechanisms. In recent years, new studies have revisited this subject with various phenomenological approaches including excited-state transition pathway calculations and *ab initio* molecular dynamics (9, 10, 34). These simulations suggest that under non-hydrostatic conditions, the quasi-body-centered cubic (bcc) oxygen sublattice that describes α-quartz can transform via a martensitic mechanism (Burgers path) to an hcp oxygen lattice (9, 42). Such calculations give credence to a model in which the oxygen scaffolding responds rapidly to compression via a diffusionless mechanism, followed by a slower reordering of the silicon cations within their newly formed environment.

Due to the fast timescales of dynamic compression experiments, it is likely that the oxygen sublattice responds readily to loading but the motion of the silicon cations is kinetically limited. A plausible model involves a shear-mediated transformation of the oxygen lattice to a quasi-hcp structure followed by the diffusion-limited reordering of silicon. Within the nanosecond timescale of shockwave experiments, kinetic barriers may prevent the silicon cations from finding the absolute energy minimum and instead local regions may minimize energy by adopting various local arrangements. The diffraction pattern from this type of partially disordered structure would contain well-defined peaks associated with an oxygen lattice, but lack the low-angle peak associated with fixed silicon order. As opposed to the fully disordered case represented by $d$-NiAs, within the structure there are local regions of short-range order, representing a distribution of periodicities centered about a $d$-spacing consistent with the observed low-angle peak. In this way, the structural transformation on the Hugoniot represents a metastable intermediate arising from a bi-modal transformation yet to reach completion.

In addition to probing the structure under compression, the time resolution of our measurements allows us to probe the structural evolution of the sample on release. Targets without a window bonded to the $SiO_2$ will have a release wave propagate back into the sample when the shock wave reaches the free surface. While the temperature can remain high as the sample releases isentropically, pressure releases rapidly when rarefication waves from the sample free surface and/or edges of the target reach the sample center. Figure 6 shows the results of one such experiment in which a novaculite sample was shocked to 35 GPa. Two XRD frames were recorded while the sample was in the partially shocked state and two XRD frames were recorded between 200-600 ns after onset of longitudinal stress release from the novaculite free surface. The impact velocity was chosen so that the peak Hugoniot pressure was below 40 GPa to avoid potentially crossing the liquidus during shock release based on the predicted isentropic release path (*15*).

In Figure 6, the reemergence of a strong peak at 9 degrees two-theta at late times is consistent with the ambient pressure α-quartz (101) diffraction peak. In addition, it appears that a portion of the sample is amorphized, evidenced by the broad feature peaked at ~8-9 degrees two-theta. While it is possible that trace amounts of the high-pressure phase are retained on release, there is no evidence for this within the resolution of our measurements. Scattering from the quenched glass is peaked at a distinctly higher Q than the FSDP for ambient fused silica, while the α-quartz (101) peak indicates that pressure is fully released. Due to overlap of the α-quartz (101) peak, amorphous scattering from the polycarbonate impactor, and amorphous scattering from the sample, it is challenging to assess the FSDP of the quenched glass with a high degree of accuracy. However, a comparison of the amorphous feature observed after release from the shocked state (~1.75 Å$^{-1}$) to the expected peak position for ambient silica glass (1.54 Å$^{-1}$), indicates the released material is consistent with a densified glass. This is evidence that the material forms a densified (diaplectic) glass during rapid stress release from the shocked state. In shock recovery experiments for Hugoniot stresses between 25-50 GPa, recovered samples show primarily amorphous material ~5-10% denser than ambient silica glass. Above a threshold pressure, where the material is expected to melt on release the recovered material corresponds to low-density glass with a structure similar to ambient fused silica. This result is consistent both with samples recovered from natural impact sites (*2*) and recent laser compression experiments that reported evidence for diaplectic glass formation in shock-compressed fused silica (*22*). In the present work, the sample undergoes a complex loading/unloading history due to multiple wave interactions and further experiments are needed to better constrain the released state.

X-ray diffraction measurements provide direct crystallographic evidence of the phase transition from α-quartz to a disordered high-pressure crystalline phase under shock compression, challenging long-standing assumptions regarding the structure of shocked quartz. Upon shock compression to 31-35 GPa, our XRD results show evidence for compressed α-quartz in combination with new XRD peaks from transformed material, indicating that the material is being shocked into a mixed-phase state. For shock pressures from 39-65 GPa, quartz fully transforms to a new structure. The compressed α-quartz densities derived from fits to the XRD patterns in the mixed-phase region are consistent with an extrapolation of the 300-K equation of state of α-quartz from static compression experiments. The diffraction data from the transformed material are distinctly different from the stishovite pattern observed for shocked fused silica. The transformed phase exhibits a broadened low-angle peak arising from silicon site disorder within a close-packed oxygen framework. This structure can be described as a defective-niccolite structure with considerable silicon short-range order. This result indicates that α-quartz transforms to a metastable high-pressure phase as opposed to crystalline stishovite, in contrast to the behavior of fused silica under shock loading. For single-crystal samples, a high-degree of crystalline texture persists above the phase transformation ruling out the amorphous intermediate proposed by molecular dynamics simulations. The disordered phase persists to the highest stresses measured, indicating that a significant kinetic barrier hinders transformation of α-quartz to stishovite on shockwave timescales. Regardless of the detailed structure of the high-pressure phase, on release the Hugoniot phase is not quenchable. From the present data, it appears that upon release the high-pressure phase reverts to the α-quartz structure in combination with amophization.

**Materials and Methods**

Time-resolved X-ray diffraction measurements coupled with gas-gun-based dynamic compression were carried out at the Dynamic Compression Sector (DCS) located at the Advanced Photon Source (APS), Argonne National Laboratory. DCS allows for the determination of the phase(s) formed under ~100-ns timescale shock loading (*48*). Starting materials consisted of synthetic single-crystal quartz and natural polycrystalline quartz. The single-crystal samples had Z-cut (001), X-cut (110), and Y-cut (100) orientations. The polycrystalline quartz was Arkansas novaculite, a nearly pure quartz rock formed by thermal metamorphism of chert. The novaculite was untextured, microcrystalline and low porosity. The measured bulk density of the novaculite was 2.640(5) g/cm$^3$. All samples were characterized at ambient conditions using powder or Laue X-ray diffraction, longitudinal sound velocity measurements and Archimedean density determination (Supplemental Material, Table S1). The novaculite samples were also characterized with Raman spectroscopy and scanning electron microscopy. All characterization results for ambient samples were consistent with literature values.

The experiments followed procedures described previously (*21, 48*). The impact configuration is shown schematically in Supplemental Material, Fig. S4. Quartz samples were cut and polished to a thickness of 1-2 mm and parallelism of better than 1mrad. For most experiments, the quartz sample was backed by a [100]-oriented single crystal lithium fluoride window. Two experiments were also performed with no window. Targets were impacted with a 10-mm diameter [100]-oriented LiF single-crystal mounted in a polycarbonate projectile. Projectiles were launched using the two-stage light-gas gun located at DCS. Projectile velocities ranged from 3.7–5.7 km/s as determined using a measurement system based on four optical beams passing through holes near the muzzle end of the gun barrel. The distance between the beams was calibrated and as the projectile passed each beam a signal change was recorded on a fast photodiode allowing the projectile velocity to be determined with a precision of better than 0.5%. For experiments on single-crystal quartz samples, Photon Doppler Velocimetry (PDV) (*49*) was also used to record the projectile velocity history until impact. Velocities measured with PDV were consistent with those recorded using the optical beam interrupts. Quartz impact stresses were calculated using impedance matching and ranged from 31-58 GPa (see Supplemental Materials). For some of the experiments an elastic shock wave propagates through the quartz followed by a slower phase transformation wave, whereas for other experiments a single shock wave brings the quartz from the ambient state directly to the peak state (*35*, *50*, *51*). The elastic shock waves have a good impedance match with the LiF windows resulting in minimal wave reflection, but when the phase transformation shock wave propagating through the quartz reflects from the LiF window, a shock wave propagates back into the sample resulting in a ~15-20% stress increase (see Supplemental Material, Fig. S1). Experimental parameters for all plate impact experiments including projectile velocities, quartz type, sample and window thicknesses, and stresses are listed in the Supplemental Material, Table S2.

X-ray diffraction data were collected in a transmission geometry such that the incoming X-ray beam made an angle of 28 degrees with the impact surface (Supplemental Material, Fig. S4). X-rays from the third harmonic of the 2.7-cm period undulator at DCS were used. The experiments were performed in the APS 24-bunch mode of operation which provides X-ray bunches of ~100 ps duration every 153.4 ns. The energy spectrum of the undulator source is peaked near 23 keV and has an asymmetric shape with a bandwidth of ~1 keV (Supplemental Material, Fig. S4). Each X-ray bunch contains ~$10^9$ photons. Use of hard X-rays with energy greater than 20 keV allowed us to use millimeter thickness quartz samples without excessive X-ray absorption in the transmission geometry. The lower-order X-ray harmonics were filtered using 250-μm thick Al and 25-μm thick Ag foils in the incident beam path, and higher X-ray harmonics were removed by reflecting/focusing the X-rays with Kirkpatrick-Baez mirrors. The typical X-ray beam size incident on the sample was ~300 μm horizontal x 800 μm vertical.

A four-frame X-ray detection system was used to record diffraction information. The X-ray detector has a 150-mm diameter active area and is positioned perpendicular to and nominally in the center of the direct X-ray beam. Before each shot, an XRD image was collected from a thin polycrystalline silicon calibration target. The sample-detector distance and the instrumental resolution function were determined by performing a Rietveld refinement (*52*) of this silicon diffraction pattern (see Supplementary Material, Fig. S2). Typical sample-to-detector distances were approximately 140 mm allowing us to record diffraction up to scattering angles of about 28 degrees.

The detector system at DCS utilizes a fast phosphor scintillator that converts scattered X-rays to visible light which is then directed to one of four intensified CCD cameras resulting in each XRD frame corresponding to a single ~100 ps X-ray diffraction snapshot. Because of the finite X-ray phosphor decay time, particularly strong diffraction peaks are sometimes retained as a ghost image in the following diffraction frame. For most experiments, diffraction images from four consecutive X-ray bunches (153.4 ns between bunches) were recorded, but X-ray bunches can also be skipped allowing the late-time released state to be examined.

The experiments were designed such that at least one X-ray diffraction frame was obtained while the phase transformation shock wave was propagating through the $SiO_2$ but before the phase transformation wave reached the rear quartz surface. These XRD frames correspond to Hugoniot states and for single-crystal quartz samples, the last such XRD frame recorded (corresponding to the most fully shocked material) was analyzed in detail to determine the structure of the shock-compressed quartz on the Hugoniot. For polycrystalline quartz samples, diffraction rings from uncompressed material ahead of the initial shock front overlap with diffraction rings from the shocked material precluding quantitative analysis of the singly shocked Hugoniot state. Therefore, analysis of polycrystalline samples was restricted to double-shock data collected after the initial phase transformation shock wave reshocks from the LiF window. A similar analysis of the reshocked state was also performed on some of the single-crystal quartz samples. For reshocked states, the latest XRD frame recorded (prior to arrival of release waves) corresponding to the most fully reshocked material was analyzed in detail to determine the structure of the reshocked quartz.

The X-ray measurement frame times relative to shockwave breakout at the rear surface of the quartz samples were determined using laser interferometry measurements either at the quartz/LiF interface or the quartz free surface. Both PDV and Velocity Interferometer System for Any Reflector (VISAR) (*53*) probes were used. The LiF windows had a vapor deposited Al mirror on the side bonded to the quartz sample. Samples without LiF windows had a vapor deposited Al mirror on the rear surface of the quartz. A VISAR probe was used to record the shock arrival time at the center of the rear surface of the quartz. Two additional PDV probes located at the same radius from sample center and co-linear with the sample center also provided shock arrival time at the rear of the quartz sample. The recorded VISAR and PDV signals were correlated to the times at which X-rays were incident on the quartz samples. The four XRD frame times relative to shock breakout are listed in Supplemental Material, Table S3.

VISAR measurements also provided particle velocity histories at the quartz/LiF interfaces (Supplemental Material, Fig. S5). The Hugoniot elastic limit (HEL) for Z-cut quartz is ~15 GPa with an elastic wave velocity of 7.52 km/s (*35*, *54*). Because of the relatively high HEL and relatively large elastic shock speeds (*35*) for Z-cut quartz, an elastic precursor is observed in all recorded wave profiles for the Z-cut quartz plate impact experiments. For other quartz crystal orientations as well as novaculite, the HEL is lower and the elastic shock velocities are also lower. The reported elastic velocities are 6.15 km/s for polycrystalline quartz (*50*) and 6.01 km/s and 6.20 km/s for X- and Y-cut quartz, respectively (*35*). This results in the elastic precursor being overdriven by the phase transformation wave for some of the higher stress experiments, consistent with previous measurements (*35*).

**Acknowledgments**

We thank Jeff Klug, Yuelin Li, Drew Rickerson, Adam Schuman, Nicholas Sinclair, Paulo Rigg and Brendan Williams of the Dynamic Compression Sector for assistance with impact experiments. Binyamin Glam and Donghoon Kim provided experimental assistance. Kellie Swadba assisted in analysis of wave-profile data. Yogendra Gupta is thanked for providing comments after a careful reading of the manuscript. The Arkansas novaculite samples were obtained from the Princeton University Mineral Collection. This research was supported by the Defense Threat Reduction Agency (HDTRA1-15-1-0048) and the National Science Foundation (EAR-1644614). Washington State University provided experimental support through the U.S. Department of Energy/ National Nuclear Security Agency Award No. DE-NA0002007. This work is based upon experiments performed at the Dynamic Compression Sector, operated by Washington State University under Department of Energy/National Nuclear Security Agency Award No. DE-NA0002442. This research used the resources of the Advanced Photon Source, a Department of Energy Office of Science User Facility operated for the Department of Energy Office of Science by Argonne National Laboratory under Contract No. DE-AC02-06CH11357.

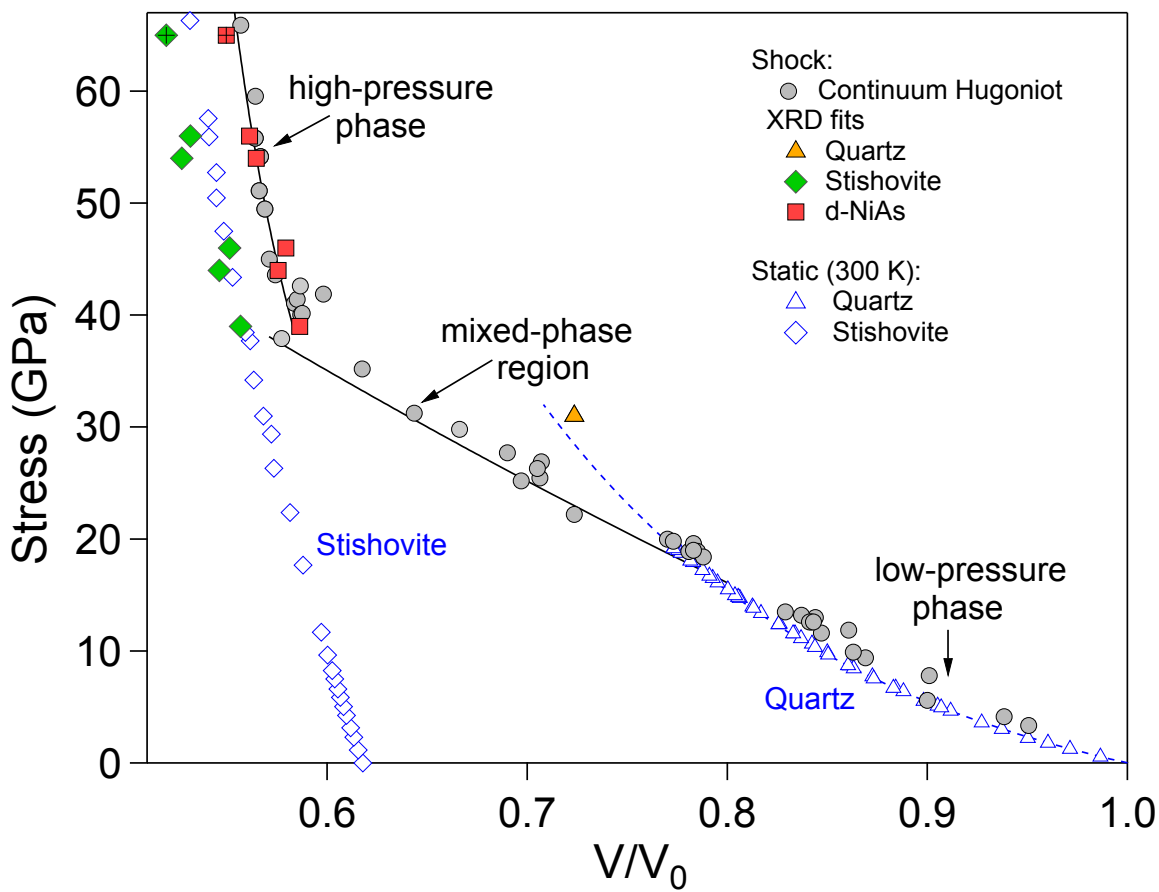

**Fig. 1**. **Shock Hugoniot curve for quartz**. Continuum Hugoniot data are shown as grey symbols (*35–38*). Black curve is a guide to the eye. Blue open symbols are 300-K static compression data for stishovite (*39*) and α-quartz (*24*). The blue dashed line is an extrapolated equation of state fit to the 300-K quartz data. Densities derived from Le Bail fits to Z-cut quartz XRD data using α-quartz, stishovite, or *d*-NiAs structures are shown as solid orange, green, and red symbols, respectively. X-ray fits to the off-Hugoniot double-shock state at 65 GPa are indicated with crosses.

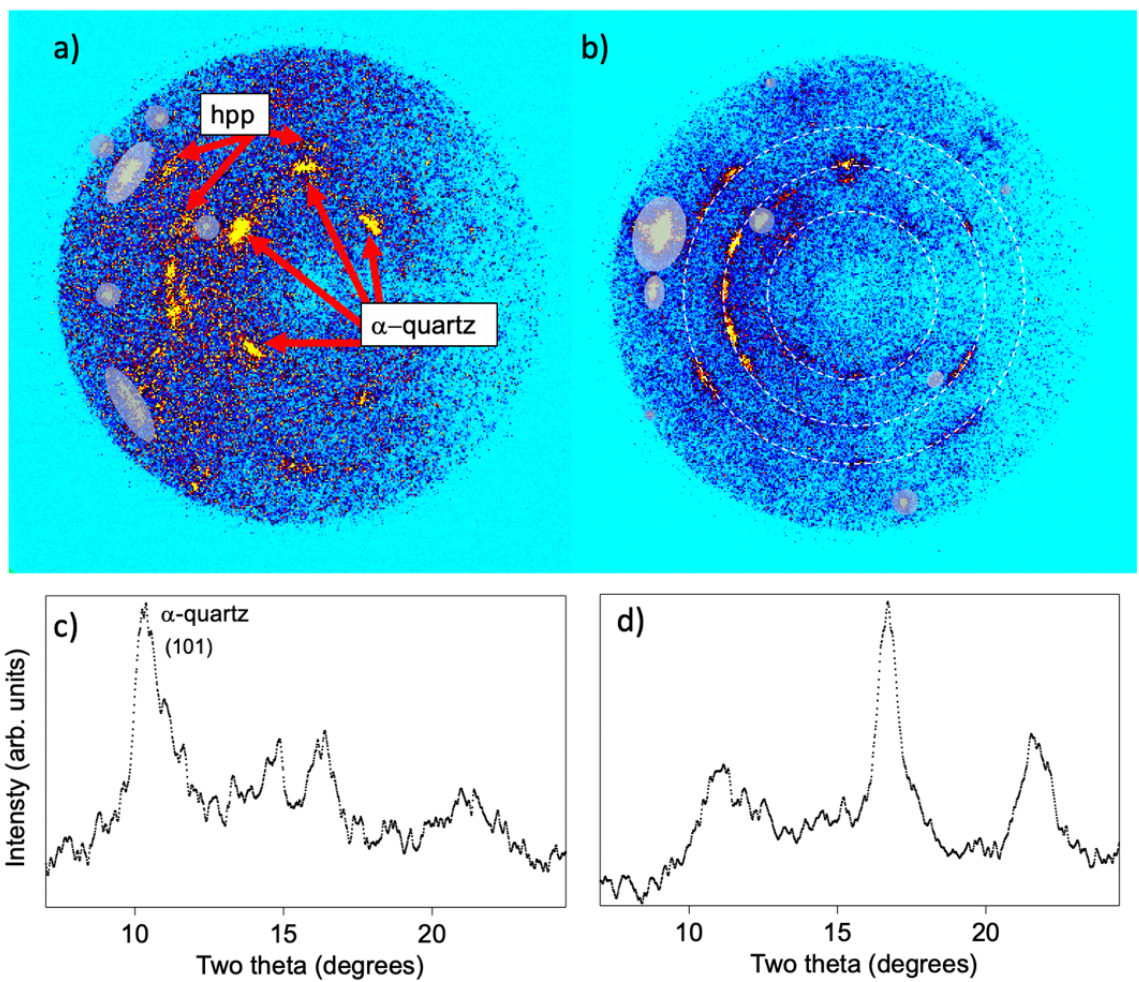

**Fig. 2. Two-dimensional X-ray diffraction images and corresponding integrated XRD patterns. a)** 2D XRD image for Z-cut quartz shock compressed to 31 GPa (shot 16-5-033, Frame 3). Diffraction is observed from both compressed α-quartz and a high-pressure phase (hpp), indicating the material has been shocked into a mixed-phase region. **b)** 2D XRD image image taken for Z-cut quartz shock compressed to ~63 GPa (shot 16-5-118, Frame 3). This frame was captured after partial reshock from the LiF window and constitutes a combination of $SiO_2$ reshocked to 67 GPa and material in the initial 58 GPa single-shock state. Dashed lines indicate three strong reflections seen in the integrated pattern (below). Grey regions in (a) and (b) are diffraction from the LiF impactor, the LiF window, or Laue spots from uncompressed or elastically compressed α-quartz; these regions were masked during azimuthal integration. Due to the diffraction geometry (Fig. S4), scattered X-rays recorded on the right half of the detector undergo higher attenuation due to greater sample absorption. **c)** Integrated XRD pattern corresponding to the 2D image shown above. Peak around ~14° is consistent with the LiF (200) peak. **d)** Integrated XRD pattern corresponding to 2D image shown above.

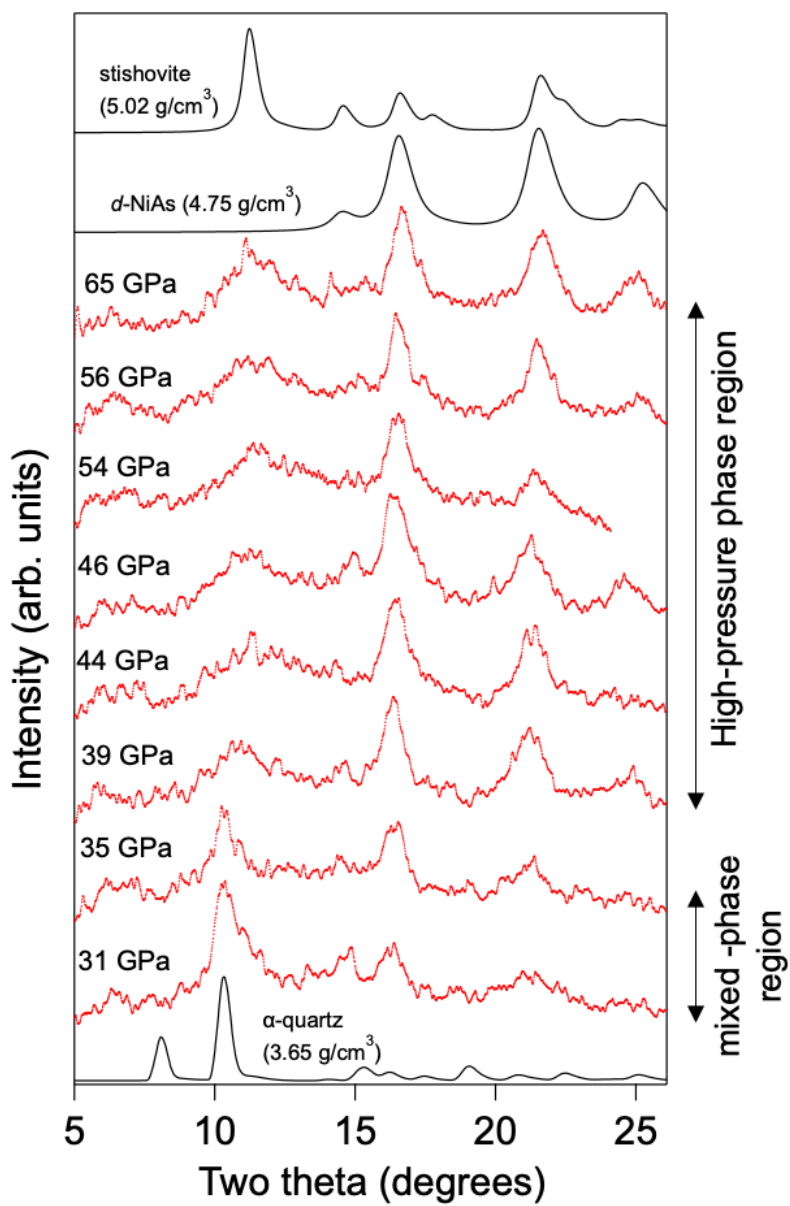

**Fig. 3. Azimuthally integrated X-ray diffraction patterns.** XRD patterns collected for a series of plate-impact experiments for Z-cut quartz starting material with peak stress states between 31 and 65 GPa. Shot numbers and X-ray frame times relative to shock breakout at the rear surface of the quartz are listed in Supplemental Material, Table S3. Note that the 54-GPa shot was collected during an experiment with a longer sample-detector distance, leading to lower angle cut-off in the data compared to the rest of the shots. Simulated diffraction patterns (accounting for spectral shape of the pink X-ray beam) are shown for compressed α-quartz (3.65 g/cm$^3$) as well as both stishovite (5.02 g/cm$^3$) and the defective niccolite structure (4.75 g/cm$^3$). The simulated patterns shown use lattice parameters based on fits to data for α-quartz in mixed-phase region and for stishovite and defective niccolite structure in high-pressure region.

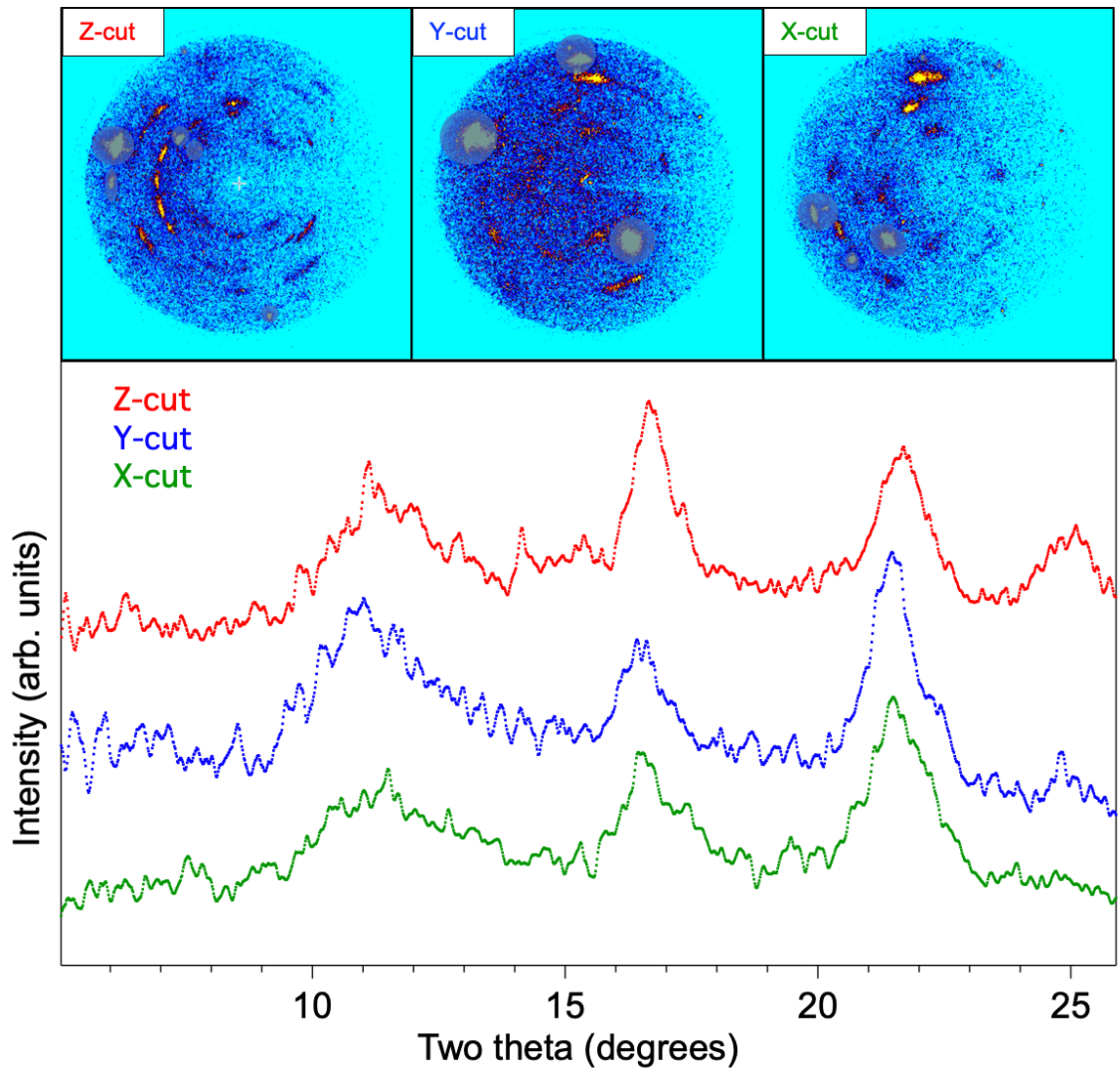

**Fig. 4**. **X-ray diffraction data for single-crystal quartz shock loaded to 56-65 GPa along different crystallographic axes**. Data shown for Z-cut (shot 16-5-022, Frame 3) and Y-cut (shot 16-5-119, Frame 3) samples were collected during double-shock experiments at 65 and 64 GPa final stress, respectively. Data for X-cut quartz (shot 16-5-126, Frame 2) were collected while the initial shock wave was propagating through the sample bringing the material to 56 GPa shock stress. Two-dimensional images are shown at the top with LiF and uncompressed quartz spots masked and integrated patterns are shown at the bottom.

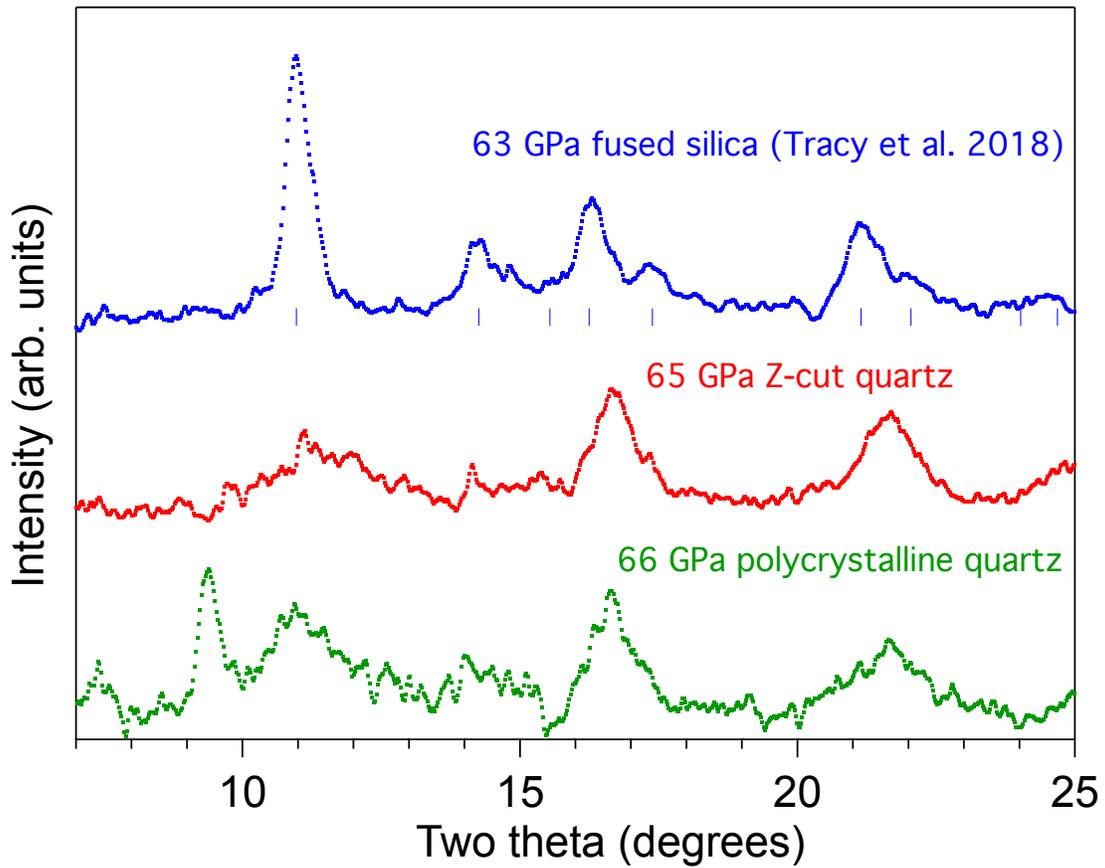

**Fig. 5. Comparison of X-ray diffraction data collected for different starting SiO₂ materials after shock compression.** Starting materials are fused silica (from experiment 17-5-037 of (*21*)), Z-cut α-quartz (experiment 16-5-022, Frame 3), and a polycrystalline natural α-quartz sample (experiment 17-5-011, Frame 3). Expected stishovite peak locations are shown as blue ticks below fused silica pattern. All patterns correspond to data collected after the initial shock wave reflects from the LiF window reshocking the sample to final stresses between 63-66 GPa.

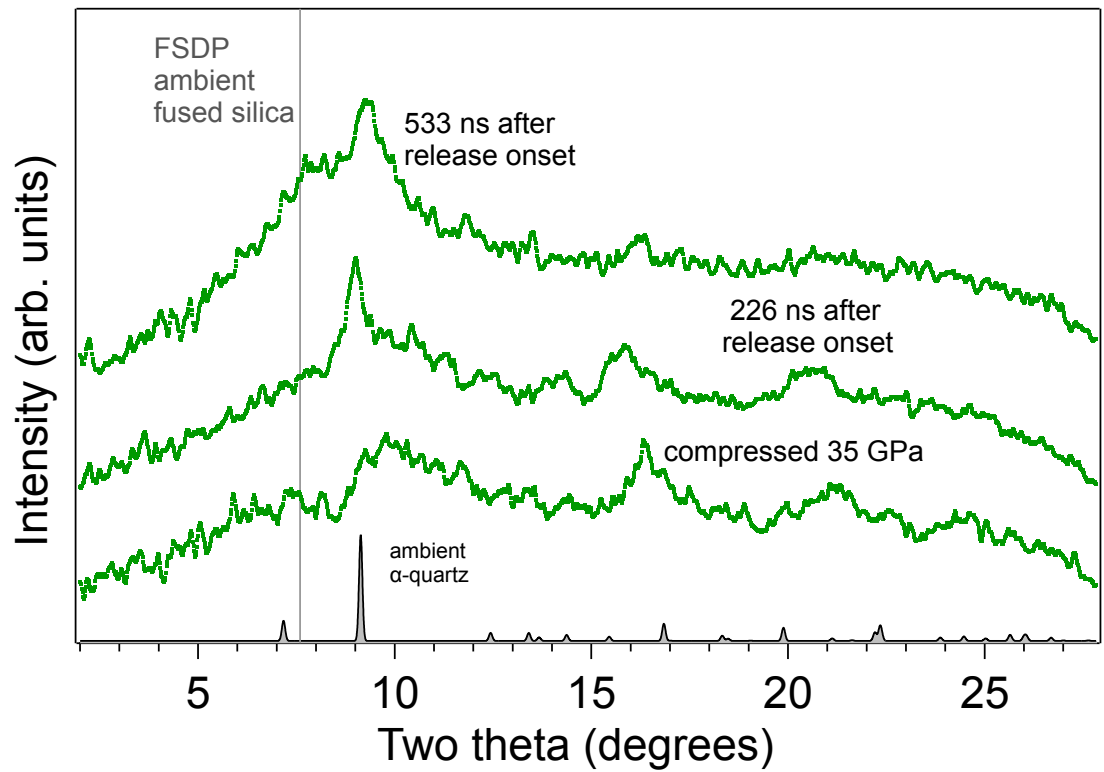

**Fig. 6**. **Integrated X-ray diffraction patterns collected after release from a peak stress of 35 GPa for novaculite starting material** (experiment 17-5-019). Compressed state data is from Frame 2 and has been background subtracted to remove uncompressed α-quartz peaks from material ahead of the shock front. Peak positions for ambient α-quartz are shown at bottom and the first strong diffraction peak (FSDP) position for silica glass at ambient pressure is shown as a vertical gray line.

**Supplementary Materials**

**Table S1. Quartz sample properties.**

|  | Density (g/cm$^3$) | Longitudinal sound speed (km/s) |
|---|---|---|
| Z-cut quartz[a] | 2.650(5) | 6.38(2) |
| X-cut quartz[b] | 2.650(5) | 5.78(1) |
| Y-cut quartz[b] | 2.650(5) | 6.03(1) |
| Novaculite | 2.640(5) | 6.0(1) |

[a]Z-cut quartz single crystal samples obtained from Boston Piezo-Optics.
[b]X-cut and Y-cut single crystal samples obtained from MTI Corp.

**Table S2. Experimental parameters for plate-impact experiments.**

| Shot Number | Quartz Sample Type/Window | Quartz Thickness (mm) | LiF Window Thickness (mm) | Projectile Velocity (km/s) | Impact Stress (GPa) | Reshock Stress[a] (GPa) |
|---|---|---|---|---|---|---|
| 16-5-118 | Z-cut/LiF | 1.266(2) | 0.749(2) | 5.66(3) | 58(1) | 67(1) |
| 16-5-022 | Z-cut/LiF | 1.267(2) | 0.738(2) | 5.52(2) | 56(1) | 65(1) |
| 16-5-007 | Z-cut/LiF | 1.259(2) | 0.750(2) | 5.38(4) | 54(1) | - |
| 16-5-023 | Z-cut/LiF | 1.249(2) | 0.740(2) | 4.9(2) | 46(3) | - |
| 16-5-029 | Z-cut/LiF | 1.247(2) | 0.706(2) | 4.71(2) | 44(1) | - |
| 16-5-120 | Z-cut/LiF | 1.267(2) | 0.697(2) | 4.41(1) | 39(1) | - |
| 16-5-127 | Z-cut/LiF | 1.268(2) | 0.705(2) | 4.01(1) | 34.5(9) | - |
| 16-5-033 | Z-cut/LiF | 1.253(2) | 0.744(2) | 3.69(3) | 31.3(8) | - |
| 16-5-119 | Y-cut/LiF | 1.038(2) | 0.681(2) | 5.47(4) | 55(1) | 64(1) |
| 16-5-126 | X-cut/none | 1.527(2) | - | 5.51(1) | 56(1) | - |
| 17-5-011 | Novaculite/LiF | 1.156(2) | 1.519(2) | 5.59(3) | 57(1) | 66(1) |
| 17-5-013 | Novaculite/LiF | 1.131(2) | 0.995(2) | 4.84(1) | 45(1) | 54(1) |
| 17-5-014 | Novaculite/LiF | 1.183(2) | 0.983(2) | 4.06(2) | 34.9(8) | 42(1) |
| 17-5-015 | Novaculite/LiF | 1.045(2) | 0.680(2) | 3.72(1) | 31.6(7) | 36(1) |
| 17-5-019 | Novaculite/none | 1.738(2) | - | 4.08(9) | 35(1) | - |

[a]Reshock stresses only listed for shots where XRD measurements from reshock were analyzed.

**Table S3. XRD frame times relative to shock breakout at rear surface of quartz.**

| Shot Number | Frame 1 time (ns) | Frame 2 time (ns) | Frame 3 time (ns) | Frame 4 time (ns) |
|---|---|---|---|---|
| 16-5-118 | -260 | -106 | 47* | 201 |
| 16-5-022 | -205 | -52* | 102* | 255 |
| 16-5-007 | -162 | -9* | 145 | 298 |
| 16-5-023 | -299 | -16 | 8* | 161 |
| 16-5-029 | -181 | -28* | 125 | 279 |
| 16-5-120 | -175 | -22* | 132 | 285 |
| 16-5-127 | -191 | -38* | 116 | 269 |
| 16-5-033 | -271 | -118 | 36* | 189 |
| 16-5-119 | -245 | -91 | 62* | 216 |
| 16-5-126 | -214 | -61* | 93 | 246 |
| 17-5-011 | -213 | -60 | 93* | 247 |
| 17-5-013 | -80 | 73* | 226 | 380 |
| 17-5-014 | -202 | -49 | 105* | 258 |
| 17-5-015 | -42 | 112* | 265 | 418 |
| 17-5-019 | -234 | -81* | 226* | 533* |

\* Denotes frames shown in main text or Fig. S7.

**Table S4. Hugoniot equation of state parameters.**

| Material | $C_0$ (km/s)[a] | $S$[a] | $u_p$ range (km/s) | Reference |
|---|---|---|---|---|
| LiF[b] | 5.201 ± 0.025 | 1.323 ± 0.009 | >0.451 | Liu et al., 2015 |
| Quartz | 1.48± 0.10<br>5.29± 0.08 | 1.80 ± 0.03<br>0.20± 0.04 | 2.46-4.55<br>1.803-2.48 | Ahrens & Johnson, 1995 |
| Polycarbonate[b] | 2.767± 0.054 | 1.249± 0.019 | 0.42-5.21 | Marsh, 1980 |

[a]$C_0$ and S are the intercept and slope of the linear shock velocity-particle velocity relationship.
[b]Ambient densities are 2.64(1) g/cm$^3$ for LiF and 1.19(1) g/cm$^3$ for Lexan.

**Shock Stress Determination**

Shock stresses were determined by impedance matching as shown in Fig. S1 for a representative experiment. The intersection of the LiF impactor Hugoniot with the quartz Hugoniot in the stress-particle velocity plane provides the impact stress. The LiF Hugoniot is determined using a linear shock velocity-particle velocity relationship as described in Table S4 (*55*). For quartz, the phenomenological shock velocity-particle velocity relations given in Table S4 (*56*) are used to calculate corresponding stress-particle velocity curves shown as solid black lines in Fig. S1; these curves provide a good match to published quartz Hugoniot states (*13, 35–38, 51, 54, 57–59*).

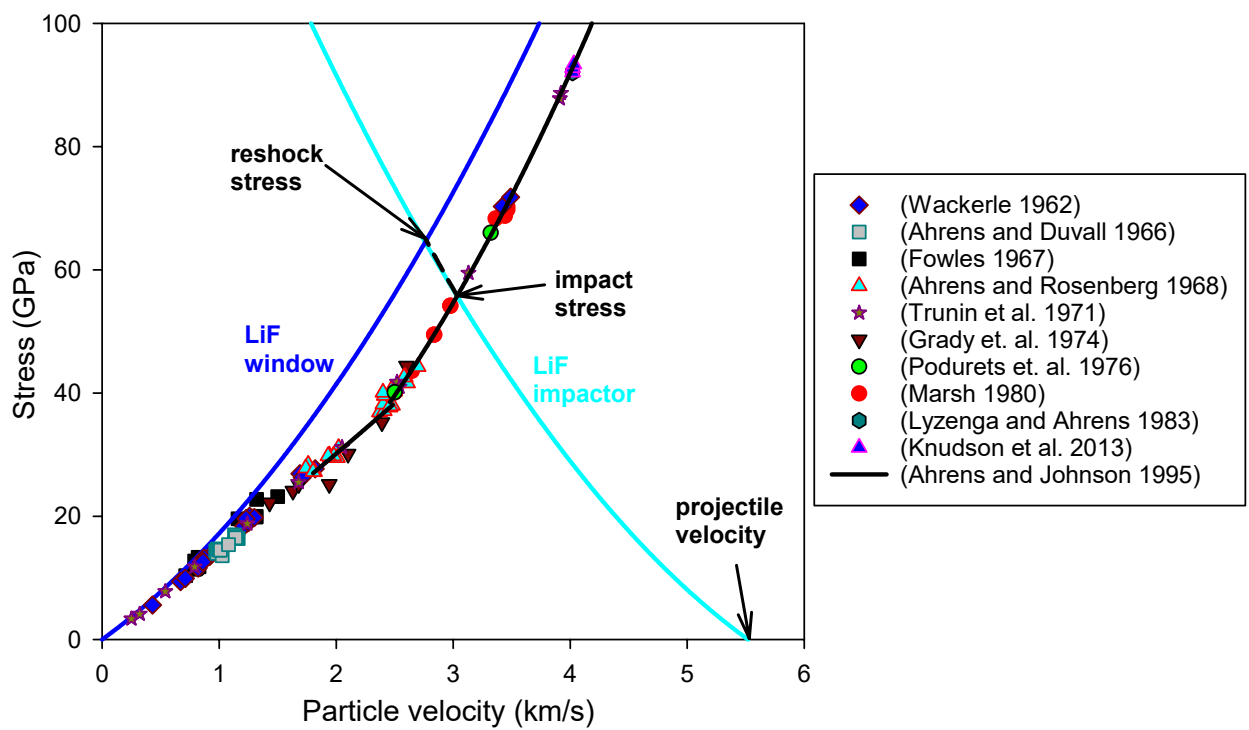

**Fig. S1. Stress-particle velocity diagram for quartz plate impact experiment 16-5-022.** The Hugoniot shock states achieved upon impact are determined by the intersection of the LiF impactor Hugoniot with an analytical form of the quartz Hugoniot (solid black lines). Symbols represent reported Hugoniot states for quartz. The reshocked state obtained after reflection of the phase transformation wave from the LiF window is also shown. The reshocked state is determined by reflecting the quartz Hugoniot around the impact state (black dashed line) and finding the intersection of this reflected Hugoniot with the LiF window Hugoniot. Calculated reshock stresses are listed in Table S2 for experiments where XRD measurements from reshocked states were analyzed.

**XRD Data Analysis**

Two-dimensional XRD images are azimuthally integrated using the program FIT2D (*60*). Bright diffraction spots from the single-crystal LiF impactor and uncompressed (or elastically compressed) α-quartz were masked prior to integration. Measured diffraction patterns exhibit characteristic peak broadening as a result of the width and asymmetry of the pink X-ray beam. The spectral flux of the incident X-rays was measured using a channel-cut Si monochrometer and a positive-intrinsic-negative (PIN) diode (see Fig. S4b). To accurately calibrate the detector and correctly determine *d*-spacings, structural refinements were performed by discretizing the measured spectral flux.

Prior to each shot, an XRD image was collected from a thin polycrystalline silicon calibration target. This data was used to determine sample-detector distance and to assess the instrumental broadening. The instrumental profile function was determined with a Rietveld refinement to the Si calibration data using the software package Maud (*52*). While the primary source of broadening is the spectral flux, additional instrumental broadening is treated with a Cagliotti function. A representative silicon calibration pattern with the corresponding fit is shown in Fig. S2.

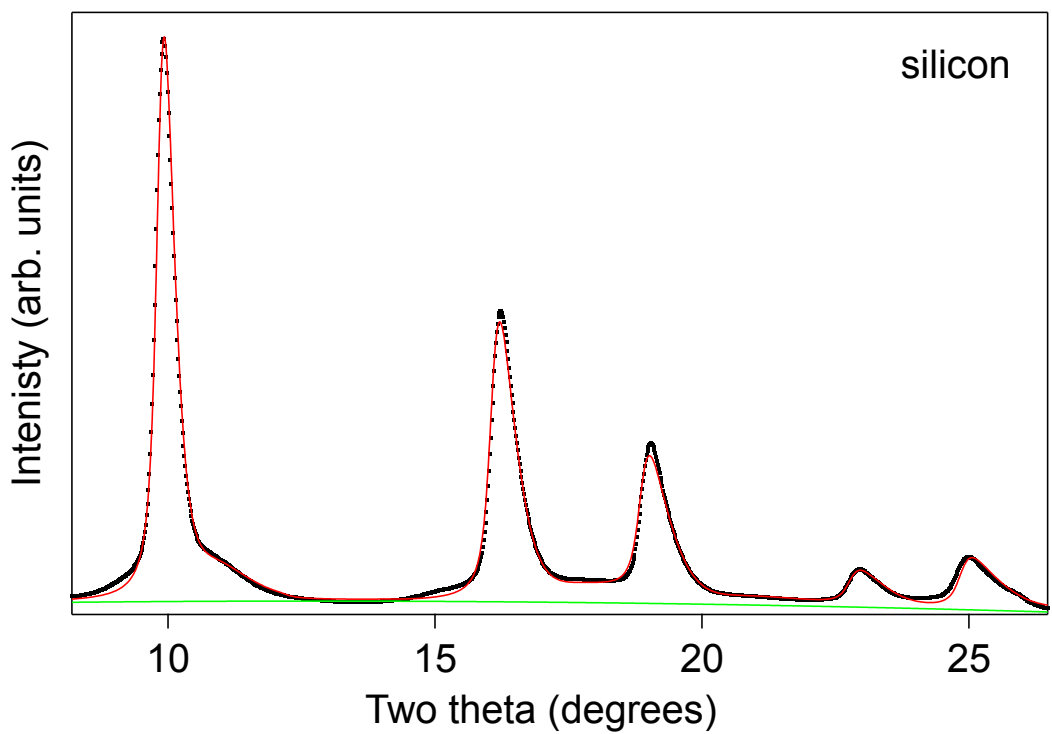

**Fig. S2. Representative Rietveld refinement for silicon calibration target.** Black curve is the measured silicon line profile. Red and green curves show the overall fit and background portion of the fit, respectively.

The sample-to-detector distance is corrected for the difference in thickness between the silicon calibration target and the quartz targets. An additional correction for target translation after impact is made using the quartz particle and shock velocities. This correction ensures the sample-detector distance for each X-ray frame reflects the distance along the beam path between the detector and the center of the shocked portion of the quartz sample.

Densities were determined for both the stishovite and *d*-NiAs structures with Le Bail fits incorporating the instrumental profile determined from the silicon calibration spectrum. The background was fit using a cubic polynomial. Modest additional broadening was incorporated to best capture the peak profiles. For fits to the *d*-NiAs structure, a Gaussian background peak was incorporated at low angle to capture the broad feature at two-theta ~12 degrees. For both phases, the unit cell volume was refined with a fixed *c/a* ratio from diamond anvil cell results (*41*, *61*). Representative fits for the two phases are shown in Fig. S3.

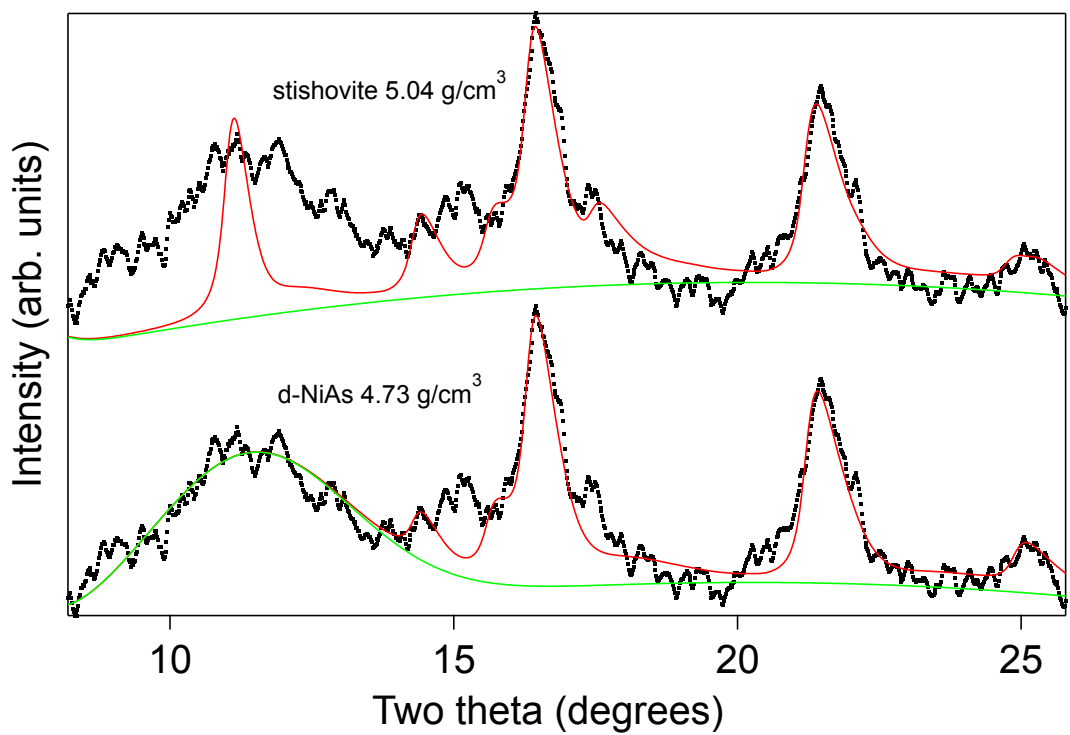

**Fig. S3. Two representative Le Bail refinements for stishovite and *d*-NiAs structures** for XRD data collected at 56 GPa (shot 16-5-022, Frame 2). Red and green curves show overall fits and the background portion of the fits, respectively. The weak peak at two-theta ~15° is consistent with the LiF (111) peak from the impactor.

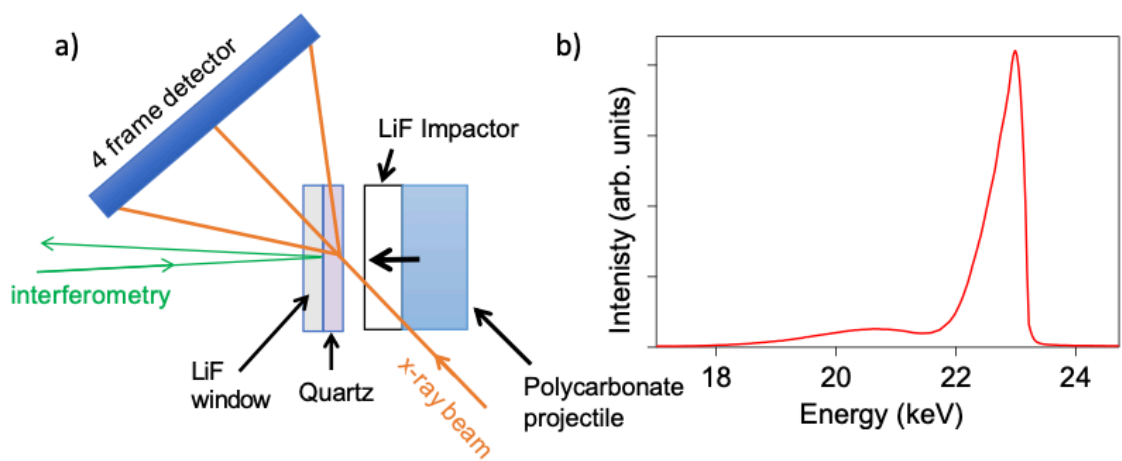

**Fig. S4. a) The experimental configuration for *in-situ* XRD measurements under shock compression.** The LiF window shown was not used in all experiments (see Table S2). **b)** Representative measured spectral X-ray flux from the 2.7-cm undulator at DCS.

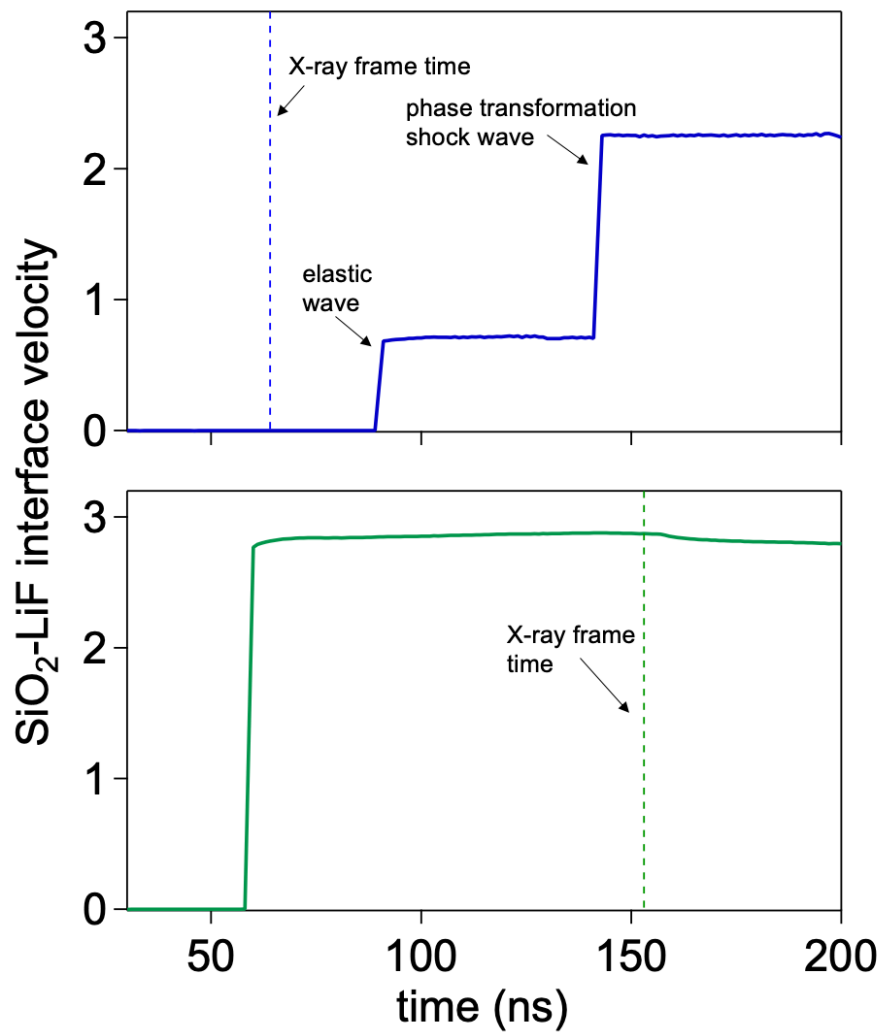

**Fig. S5. Representative SiO$_2$-LiF interface velocity histories** for Z-cut quartz shocked to 44 GPa (blue, shot 16-5-029) and novaculite shocked to 57 GPa (green, shot 17-5-011). A two-wave structure consisting of an elastic precursor and a phase transformation wave is observed for Z-cut quartz, but the elastic precursor is overdriven by the phase transformation wave in the case of the polycrystalline novaculite sample. The vertical dashed lines show the times at which the analyzed XRD frames were recorded.

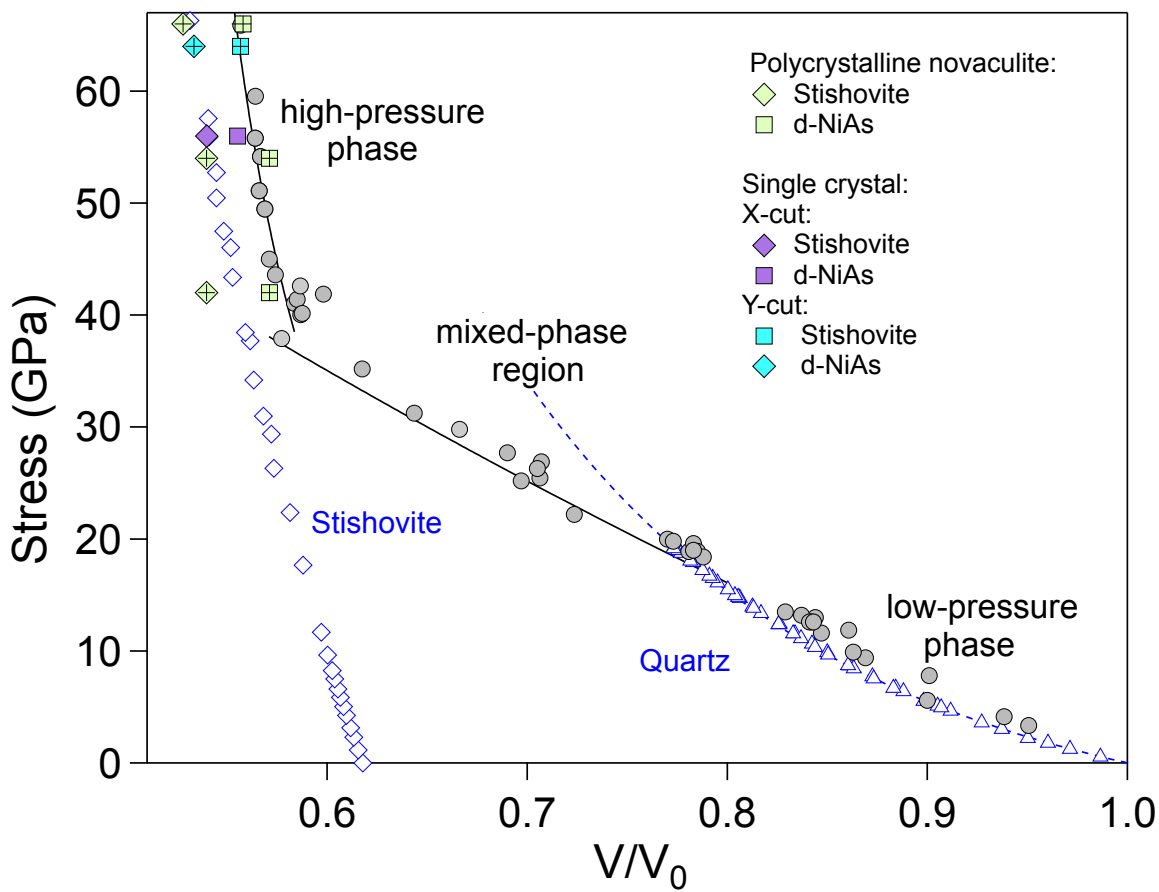

**Fig. S6. Shock Hugoniot data for quartz.** Continuum Hugoniot data are shown as grey symbols (*35–38*). Black curve is a guide to the eye. Blue open symbols are 300-K static compression data for stishovite (*39*) and α-quartz (*24*). The blue dashed line is an extrapolated equation of state fit to the 300-K quartz data. Densities derived from fits to polycrystalline novaculite, Y-cut quartz, and X-cut quartz are shown as solid green, blue, and purple symbols, respectively. X-ray fits to off-Hugoniot ring-up shots are indicated with crosses.

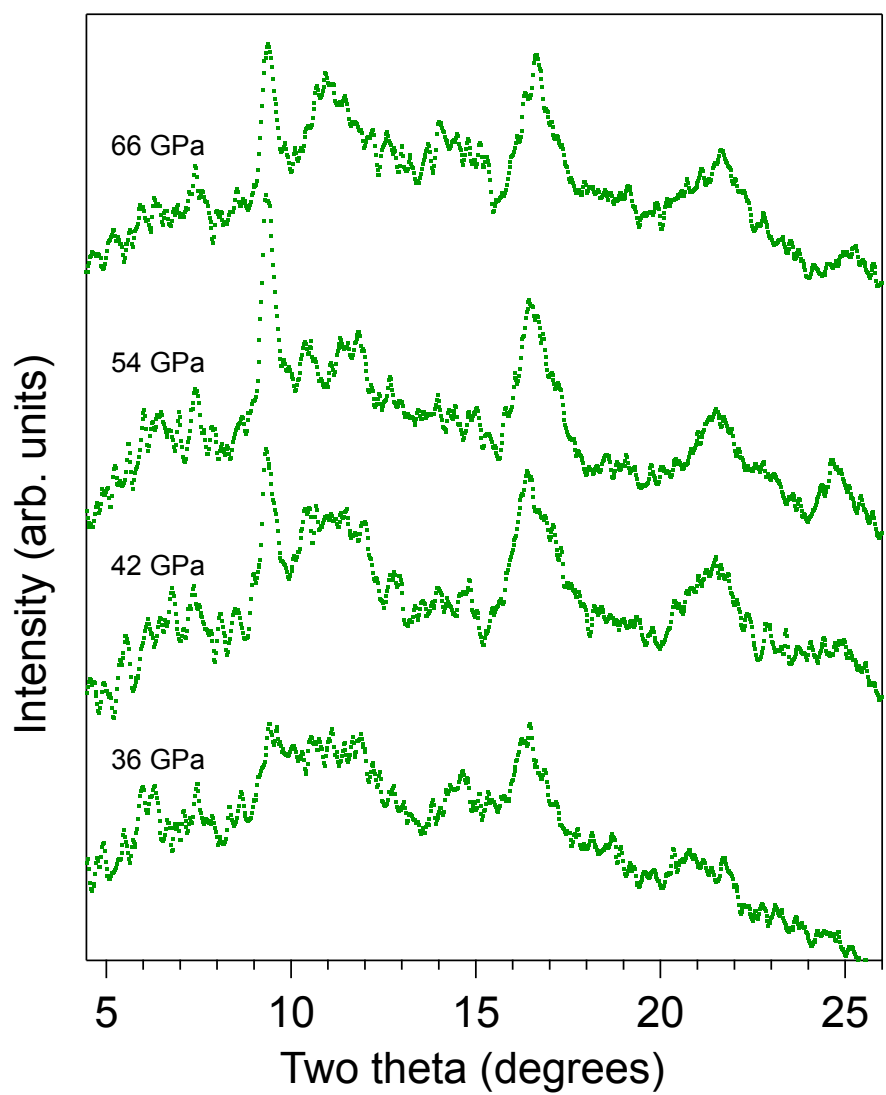

**Fig. S7. Azimuthally integrated X-ray diffraction data collected for a series of plate-impact experiments for polycrystalline novaculite** starting material with peak stress states between 36 and 66 GPa. The sharp peak at 9.5 degrees corresponds to a ghost peak from the ambient strong (101) α-quartz reflection. All results shown correspond to ring-up measurements, after reshock from the LiF window such that >85% of $SiO_2$ is in a reshocked state.

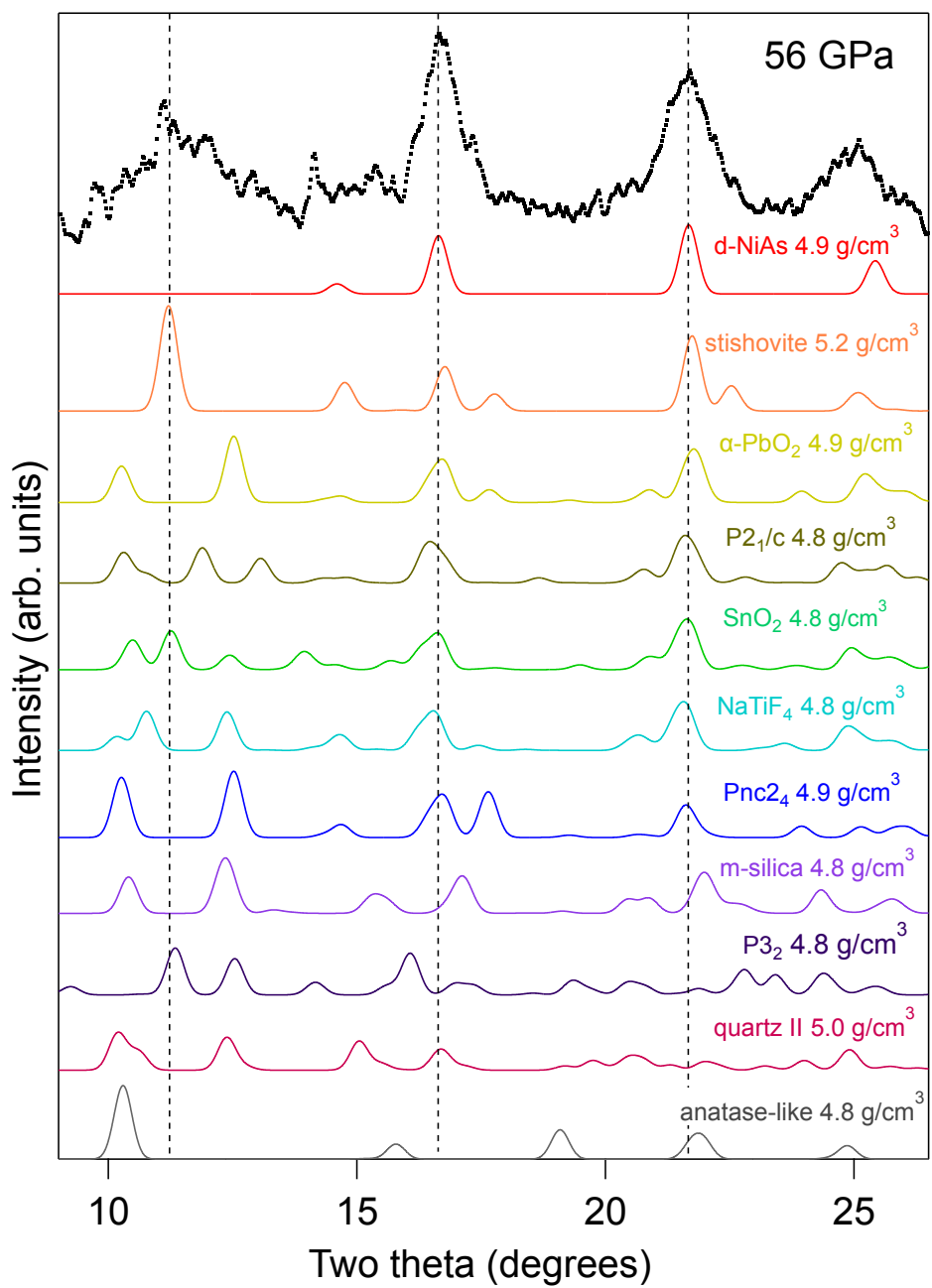

**Fig. S8. Comparison of simulated diffraction patterns for candidate high-pressure SiO₂ structures** from theory and experiment (*5–7, 10, 41, 62*) with measured Z-cut quartz XRD line profile at 56 GPa (shot 16-5-022, Frame 2).

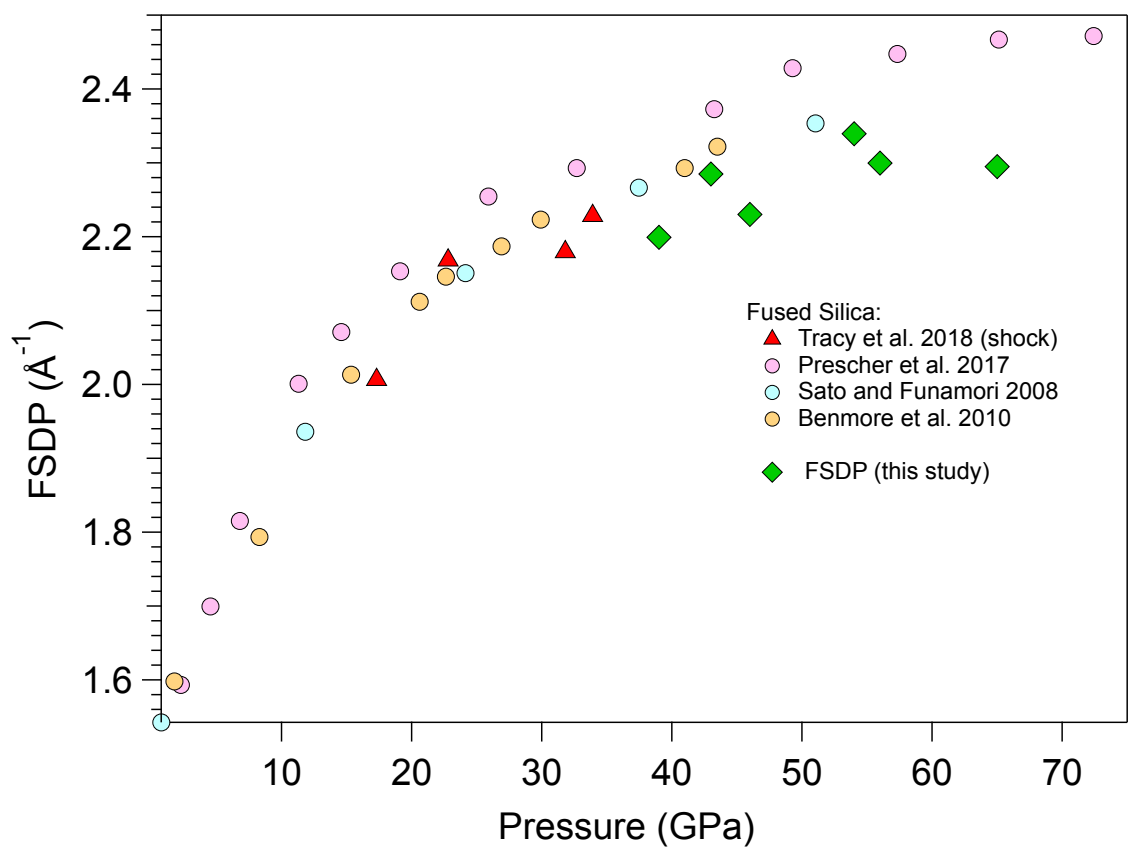

**Fig. S9. Comparison of the position of the first sharp diffraction peak (FSDP) for fused silica** starting material from shock and static studies (*21, 29–31*) compared to present results for the low-angle peak from Z-cut quartz shots (green diamonds).